# Robust orbital-angular-momentum-based underwater acoustic communication with dynamic modal decomposition method

Liulin Li,[1] Bingyi Liu,[1,a)] and Zhongyi Guo,[1,a)]

[1] *School of Computer Science and Information Engineering, Hefei University of Technology, City, Hefei 230009, China*

Recently, acoustic communication employing Orbital Angular Momentum (OAM) opens another avenue for efficient data transmission in aquatic environments. Current topological charge (TC) detection of OAM beams relies on the orthogonality among different-order OAM beams. Such strategy requires data collection from the entire acoustic field, which inevitably reduces the efficiency and increases the bit error rate (BER). To address these challenges, this study proposes a modified Dynamic Modal Decomposition (DMD) method by partially sampling the acoustic field for precise TC detection. Numerical simulations confirm the accuracy of this approach in extracting single or multiple TCs magnitudes within a partially-sampled acoustic field. We theoretically compare the performance of the modified DMD approach with conventional orthogonal decoding method. Simulation results indicate that our modified DMD scheme exhibits lower BER under the same noise interference and is more robust to the array misalignment. This research introduces an efficient demodulation solution for acoustic OAM communication, offering potential benefits for simplifying receiver array design and enhancing long-distance underwater data transmission.

[a] Email: bingyiliu@hfut.edu.cn
[a] Email: guozhongyi@hfut.edu.cn

## I. INTRODUCTION

In the current landscape of communication, the research on signal modulation and multiplexing techniques primarily revolves around the fundamental wave characteristics like amplitude, phase, frequency, and time, which aims at improving the overall capacity of communication systems by leveraging the available attributes of waves. Besides the conventional signal modulation methods like amplitude modulation[1-3] and phase-shift keying,[4,5] signal multiplexing techniques are widely employed as well, such as frequency-division multiplexing technology,[6-7] time-division multiplexing technology,[8,9] and wavelength-division multiplexing technology,[10,11] etc. However, the persistent exploration of available attributes of waves has nearly reached an inherent limit. Hence, there exists an exigent requirement to figure out another dimension for communication. Spiral beams, those that are characterized by helical phase fronts, show phase singularities at their centers, and possess a donut shape intensity profile. The phase front of spiral beam is mathematically given by $\exp(il\varphi)$, where $l$ refers to the topological charge (TC), and $\varphi$ is the azimuthal angle.[12] Notably, spiral beams inherently carry the orbital angular momentum (OAM) as an additional degree of freedom, which shows great potential to address the current constraints on communication resource exploitation.

In recent years, optical vortex (OV) has experienced rapid development in the field of wireless communications, yielding numerous achievements in the generation, transmission, and detection of OVs.[13-16] In 2004, Gibson et al.[17] established a 15-meter optical communication link based on the OAM shift keying (OAM-SK) mechanism. They employed holograms to transform the incident light beam into eight distinct OAM modes. At the receiving end, a two-dimensional fork grating hologram was utilized for demodulation, which successfully realized data transmission and decoding. In 2014, Krenn et al. conducted a communication experiment spanning a distance of 3 kilometers over Vienna city to demonstrate the feasibility of long-distance OAM-SK communication.[18] They collected intensity patterns at the receiver and employed machine learning technique for information demodulation.

Furthermore, the spatial dimension of OAM beams can be combined with other information dimensions such as amplitude,[19] wavelength,[20-22] and polarization,[23,24] resulting in tremendous increases in the channel capacity. In 2016, Zhu et al.[23] utilized a coding and decoding scheme that combined three linear polarization modes with one OAM mode for image transmission over kilometer-scale optical fibers, achieving zero-error data transmission. In 2018, Fu et al.[19] introduced a modulation approach that separately manipulated the spatial dimensions of OAM mode and amplitude. This strategy allowed for the transmission of more information with fewer OAM modes, which efficiently enhances the communication capacity.

Due to the heightened absorption of water molecules within the microwave and mid-to-far-infrared spectrum, as well as the proclivity of optical waves to be obstructed and scattered by minuscule marine particles, acoustic waves offer a distinct advantage over electromagnetic waves in underwater communication. Research in the field of acoustic communication employing OAM has yielded substantial achievements.[25-28] Marston et al.[29] firstly proposed data transmission through the manipulation of the TC of an acoustic vortex (AV) beam, where rapid changes in helicity were induced by phase modulation of different quadrant transducers.[30] It only requires a four-element array[31] as a receiver to capture underwater acoustic signals without the need for additional data processing to reconstruct the TC carried by the AV beam. While this measurement method achieves only $l = +1$ or $l = -1$ decoding, it presents valuable insights for the application of acoustic OAM communication. In 2017, Shi et al.[32] introduced a high-speed underwater acoustic communication technology rooted in OAM principles, which yielded an impressive data transmission rate of 8±0.4 (bit/s)/Hz by utilizing eight distinct OAM modes of TCs ranging from −4 to +4. In 2018, Jiang et al.[33] pursued a different avenue, i.e., acoustic metamaterials, to directly manipulate the AV beams carrying different TCs. Notably, this methodology differs from the conventional techniques that encode information onto OAM's switch states. It was observed that OAM beams with various TCs not only exhibited mutual orthogonality but were also independent of signal amplitude and phase. Consequently, OAM multiplexing technology seamlessly integrated with existing multiplexing techniques, opening another communication dimension without incurring additional channel overhead. In 2019, Zhang et al.[34] harnessed AV multiplexing technology to achieve the transmission of standard grayscale images containing 256×256 pixels. At the receiving end, a square receiver array was employed to reconstruct these images. In this endeavor, each pixel was encoded in an On-Off Keying (OOK) format using OAM acoustic carriers, resulting in a spectral efficiency of 8 (bit/s)/Hz, an eightfold improvement in information transmission rates compared to the direct OOK modulation. In the pursuit of structurally straightforward and readily controllable transmitter-receiver array designs, Li et al. in 2020 established a theoretical model for OAM beam communication by using a single-ring transmitter-receiver array.[35] Subsequently, they substantiated this model by implementing an experimental system featuring a single-ring array comprising 16 acoustic sources. Continuing this trajectory of advancements, Guo et al.[36] in 2020 further enriched the field by employing a single-ring receiver array and utilizing spectral decomposition theory to achieve the demodulation of multiplexed vortex acoustic fields. Meanwhile, in our previous work, it has been demonstrated that measuring the TC by using the interference or diffraction principles of AV beams is feasible. We employed circular aperture array interferometric screens,[37] annular triangular apertures, and elliptical apertures,[38] respectively, to achieve the detection of AVs.

However, current methodologies for extracting specific TCs from multiplexed AVs fields predominantly fall into two categories. The first method involves using acoustic metamaterials to manipulate the TCs of multiplexed AV beams. When these AV beams are selectively transformed into plane waves, it becomes possible to detect the information carried by different OAM channels with only one receiver. However, acoustic metamaterials of complicated configurations inevitably introduce energy loss and reverberation, making it less suitable for practical applications. The second approach involves using a two-dimensional sampling array or a complete circular sampling array to collect comprehensive acoustic field data. Decoding formulas, based on the orthogonality among different modal AV beams, are then constructed to extract the OAM modes. However, two-dimensional data collection often requires numerous sampling points, and an inadequate number of sampling points could significantly influence decoding results. Due to the inherent divergence characteristic of spiral sound beams, i.e., the vortex radius gradually expands as the propagation distance increases, it often becomes impractical due to the excessively large size of the entire receiver array. Especially in practical applications, when there are obstacles along the propagation path, it is often impossible to obtain the distribution of the entire acoustic field. To address these challenges, in this study, we introduce a Dynamic Mode Decomposition (DMD) approach. DMD is initially proposed in fluid dynamics for calculating the oscillatory frequencies and decay/growth rates of spatiotemporal signals,[39] then it is applied to signal decomposition[40] and reconstruction.[41] In 2019, Zhang et al.[42] introduced DMD into the realm of electromagnetic vortex fields, replacing spatiotemporal signals evolving in space and time with spiral electric field signals. They demonstrated that the DMD method, in contrast to the Fourier Transform (FT)-based OAM mode decomposition, not only extracts the OAM indices but also extracts the amplitudes of each OAM mode. Moreover, they illustrated the accurate reconstruction of the original field's OAM indices and the corresponding amplitudes by merely using local electric field information. This approach achieves TC extraction by collecting partial information of vortex field, which offers valuable insights for other OAM-based wireless communications. Consequently, we endeavor to introduce this approach into the field of acoustic vortex communication.

In this paper, we propose a modified DMD approach to extract the TCs carried by the multiplexed AV field. By replacing the time-domain information with azimuthal information from the AV field, we can extract the TC $l$ rather than angular frequency $\omega$. Subsequently, we demonstrate the extraction of both single and multiple TCs in the acoustic field, which is realized by rearranging the acoustic field data from a circular arc-shaped receiver array. This approach avoids the limitation associated with traditional orthogonality-based decoding approaches. Upon comparison with conventional orthogonal decoding methods, our modified DMD approach exhibits enhanced robustness in the presence of noise interference, receiver-transmitter array misalignment, or flipping, and various other specialized scenarios. This increased resilience positions the DMD approach for wide-ranging applications in the field of acoustic communication.

## II. PRINCIPLE AND METHOD

Compared to optical and electromagnetic waves, acoustic waves offer unique advantages in underwater communication. Figure 1(a) depicts a schematic diagram of an underwater AV communication system. The employment of circular phased arrays, which is a convenient method to generate vortex beams, has found extensive applications in the fields of electromagnetics and acoustics.[35] Notably, a single circular emitting array can concurrently generate multiple AV beams with different TCs. As depicted in the right panel of Fig. 1(b), the circular phased array employed in this study comprises $M$ transducers, which are uniformly distributed along a circular perimeter with a radius of $a$. The angular separation between two adjacent sound sources is denoted as $\Delta \varphi = 2\pi / M$, then the position of the $m$-th acoustic source in cylindrical coordinate system is expressed as $(a, \varphi_m, 0)$, where $\varphi_m = 2\pi(m-1)/M$. To generate AV beam of TC $l$, it is essential to set the initial phase of the $m$-th transducer as $\phi_m = \left[2\pi l(m-1)\right]/M$. Based on the principle of point source propagation theory, the acoustic pressure $p(r, \varphi, z, t)$ produced by the $m$-th acoustic source is

$$p(r,\varphi,z,t) = \frac{A_0 \exp(i\phi_m)}{D} \exp(i\omega t)\exp(-ikD), \tag{1}$$

where $D = \sqrt{(r\cos\varphi - a\cos\varphi_m)^2 + (r\sin\varphi - a\sin\varphi_m)^2 + z^2}$ denotes the distance between an arbitrary observation point on the observation plane $(r, \varphi, z)$ and the position of the $m$-th sound source $(a, \varphi_m, 0)$. And $A_0$ represents the acoustic pressure amplitude, $t$ denotes the time, $k = \omega/c$ represents the wave number, $c$ is the speed of sound, $\omega = 2\pi f$ is the angular frequency of the acoustic wave, $f$ is the operating frequency. By superimposing the signals radiated from the $M$ sound sources, the acoustic pressure at location $(r, \varphi, z)$ can be obtained as

$$p(r,\varphi,z,t) = \sum_{m=1}^{M} \frac{A_0 \exp(i\phi_m)}{D} \exp(i\omega t)\exp(-ikD). \tag{2}$$

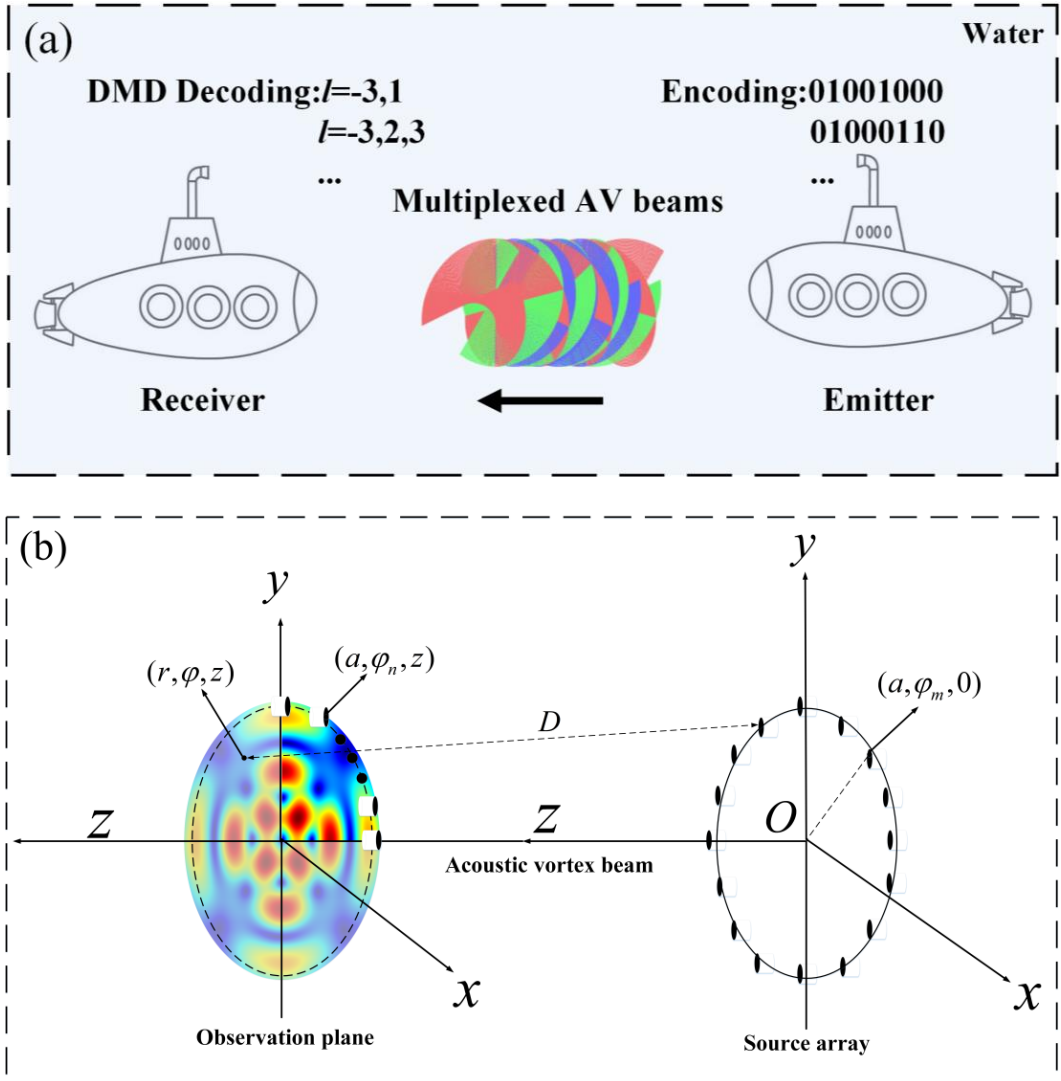

FIG. 1. (Color online) (a) Schematic of underwater AV communication system. (b) Schematic diagram of DMD demodulation based on arc-receiver array.

By precisely controlling the phase of each sound source, it is possible to generate multiple coherently superimposed AV beams with different TCs. The resultant acoustic field from the superposition of multiple distinct TCs can be expressed as

$$p_{mul}(r,\varphi,z,t) = \sum_{s=1}^{S}\sum_{m=1}^{M} \frac{A_0 \exp(i\phi_{s_m})}{D} \exp(i\omega t)\exp(-ikD), \quad (3)$$

where $\phi_{s_m} = [2\pi l_s(m-1)]/M$, with $l_s$ denoting the TC of $s$-th AV beam and $S$ represents the total number of TCs in the acoustic field. Due to the limitations imposed by the number of sound sources, the maximum attainable TC produced by $M$ sound sources is $l_{max} = fix[(M-1)/2]$, where $fix$ represents the floor function.[43] According to the unique distribution characteristics of the AV beam, the acoustic pressure distribution of multiplexed AV beams can be simplified as

$$p_{mul}(r,\varphi,z,t) = \sum_{s=1}^{S} A(l_s,r,z)\exp(il_s\varphi)\exp(i\omega t). \quad (4)$$

The items $A(l_s,r,z)$ and $\exp(il_s\varphi)$ represent the radial acoustic pressure distribution and helical phase distribution of a single AV beam with $l = l_s$ at a distance $z$ from the sound source. Previous research has revealed that the vortex radius of AV beams increases with the growing TC. Furthermore, as the propagation distance increases, the vortex radius expands, exacerbating the beam's divergence effects.[43,44] Without the aid of specialized structures, the current approach for extracting multiple TCs within a multiplexed AV beams relies predominantly on the orthogonality among AV beams with different TCs.[34,35] This method often requires a substantial number of receiving probes organized into a two-dimensional sampling array at the receiving end to collect acoustic field data, where the decoding effectiveness is significantly influenced by the number of sampling points. The approach suggested by Li employs a single-ring receiving array comprising 16 receiving transducers,[35] which significantly simplifies the design of the receiving array compared to traditional two-dimensional arrays. Nevertheless, with increasing propagation distance, the divergence of AV beams intensifies, causing a substantial increase in the radius of the single-ring receiving array. To maintain decoding accuracy, the receiving array requires a greater number of elements. These factors collectively pose challenges to practical applications. To address these issues, we propose a novel approach that employs a circular arc-shaped receiving array for the collection of acoustic field data. This approach relies on a modified DMD technique to extract specific topological charges from the multiplexed AV beams. As depicted in the left panel of Fig. 1(b), $N$ receiving transducers are evenly positioned on a 1/4 circular arc of radius $a$. The spatial position of the $n$-th receiving transducer can be expressed as $(a, \varphi_n, z)$, with an angular separation of $\Delta\varphi' = \pi / [2(N-1)]$ between adjacent transducers. By rearranging the data collected with the receiving array, we obtain two data matrices

$$\begin{cases} P = [p(a, \varphi_1, z), p(a, \varphi_2, z) \cdots p(a, \varphi_{N-1}, z)] \\ P' = [p(a, \varphi_2, z), p(a, \varphi_3, z) \cdots p(a, \varphi_N, z)] \end{cases}, \quad (5)$$

where $p(a, \varphi_n, z)$ represents the acoustic pressure detected by the $n$-th receiving transducer. Since the data matrix in the above equation consists of only one row, then the DMD approach allows us to extract an individual TC. However, when the acoustic field comprises multiple TCs, we need to alter the arrangement of the data matrix as follows

$$P = \begin{bmatrix} p(a, \varphi_1, z), p(a, \varphi_2, z) \cdots p(a, \varphi_{N-g}, z) \\ p(a, \varphi_2, z), p(a, \varphi_3, z) \cdots p(a, \varphi_{N-g+1}, z) \\ \vdots \\ p(a, \varphi_g, z), p(a, \varphi_{g+1}, z) \cdots p(a, \varphi_{N-1}, z) \end{bmatrix}, \quad (6)$$

$$P' = \begin{bmatrix} p(a, \varphi_2, z), p(a, \varphi_3, z) \cdots p(a, \varphi_{N-g+1}, z) \\ p(a, \varphi_3, z), p(a, \varphi_4, z) \cdots p(a, \varphi_{N-g+2}, z) \\ \vdots \\ p(a, \varphi_{g+1}, z), p(a, \varphi_{g+2}, z) \cdots p(a, \varphi_N, z) \end{bmatrix}. \quad (7)$$

By rearranging the data matrix as presented in the above equations, we can extract up to $g$ distinct modes. This enables the extraction of various distinct TCs from the multiplexed AV beams. Following this, by leveraging the linear relationships between consecutive sample points, we concatenate the two data matrices with a matrix $A$ as follows

$$P' = AP. \tag{8}$$

Matrix $A$ is the system matrix that encompasses numerous dynamic relationships. The eigenvalues and eigenvectors of $A$ serve as indicators of the system's dynamic characteristics. The DMD method involves mathematical transformations of the data matrix to extract the principal eigenvalues and eigenvectors of matrix $A$. In this study, we intend to employ a combined approach that integrates Singular Value Decomposition (SVD) and Proper Orthogonal Decomposition (POD) to achieve DMD. Initially, SVD is used to construct the similarity matrix $F_{DMD}$ for matrix $A$. By applying SVD to matrix $P$, we obtain

$$P = U\Sigma V^*, \tag{9}$$

$U$ and $V$ respectively represent the left singular vector matrix and the right singular vector matrix, the symbol * denotes the conjugate transpose. Substituting the Eq. (9) into the Eq. (8) can yield

$$\begin{cases} P' = AU\Sigma V^* \\ U^*P'V\Sigma^{-1} = U^*AU = F_{DMD} \end{cases} \tag{10}$$

Therefore, matrix $F_{DMD}$ can be used to approximate matrix $A$. Subsequently, by performing an eigenvalue decomposition on matrix $F_{DMD}$, we have

$$F_{DMD}W_j = \mu_j W_j, \tag{11}$$

where $\mu_j$ and $W_j$ represent the eigenvalues and eigenvectors of the matrix $F_{DMD}$, respectively. Therefore, we can obtain the eigenvector representation of the $j$-th DMD mode as $\phi_j = UW_j$,[45] and since there exists a spatial angle difference $\Delta\varphi'$ between $P$ and $P'$, the eigenvalue $\omega_j$ of the $j$-th DMD mode can be expressed as

$$\omega_j = \omega_j^r + i\omega_j^i = \ln(\mu_j)/\Delta\varphi'. \tag{12}$$

The real and imaginary parts of $\omega_j$ denote the growth/decay rate and frequency value of the associated DMD mode.[45] In this context, they respectively represent the angular acoustic pressure growth/decay rate and the TC of the corresponding AV beam. Utilizing the methods outlined above, with the reconstruction of the acoustic field using $g$ modes, the acoustic field information collected at the $n$-th receiver transducer is expressed as follows

$$p(a,\varphi_n,z) = \sum_{j=1}^{J}\phi_j e^{\omega_j\varphi_n}b_j = \sum_{j=1}^{J}\phi_j e^{\omega_j^r\varphi_n}e^{i\omega_j^i\varphi_n}b_j, \tag{13}$$

$b_j$ represents the amplitude of the $j$-th mode, as evident from the above equation, by rearranging the data matrix, we extract $J$ modes, $e^{\omega_j\varphi_n}$ and $e^{il\varphi}$ are closely related. Due to the circular acoustic pressure distribution of AV beams, the acoustic pressure remains constant at different angular positions at the same radius. Therefore, modes with a non-zero real part of $\omega_j$ are considered as evanescent waves that can be discarded. Only when the real part of $\omega_j$ is zero does the imaginary part $\omega_j^i$ of $\omega_j$ correspond to a correct OAM mode.

## III.    NUMERICAL STUDIES

In numerical studies, simulations are performed in an underwater environment with $c_0$=1500 m/s and $\rho_0$=1000 kg/m$^3$. A circular phased array consisting of 16 transducers operating at a carrier frequency $f_0$=50 kHz is used as the source. The array has a radius $a$=10 cm, and the angular separation between adjacent transducers is $\Delta\varphi = \pi/8$. By adjusting the phase difference of $2\pi l/16$ between neighboring transducers' drive signals, the circular phased array can emit an AV beam with TC $l$. Finally, data collection for the acoustic field is performed on an observation plane measuring 40×40 cm, located at a distance of $z$= 40 cm from the source plane.

Firstly, we conducted numerical simulations on acoustic fields of AV beams with $l$=1-4 by Eq. (1). The corresponding cross-sectional pressure and phase distributions are plotted in Figs. 2(a1)-(d1) and Figs. 2(a2)-(d2) respectively. It is evident that AV beams exhibit a distinctive annular acoustic pressure distribution. Subsequently, we proceeded to collect acoustic field data by configuring the receiving array with a radius of $r$=10 cm, employing 16 receiving transducers uniformly positioned along a quarter-circle arc. The angle between adjacent receiving transducers is also $\pi/32$. Given that the acoustic field contains only one mode, we only need to set $g$=1. Following the methodology discussed earlier, we restructured the acquired acoustic field data to form two data matrices with dimensions of 1×15, and utilized the DMD technique to extract a single eigenvalue from the mapping matrix $A$.

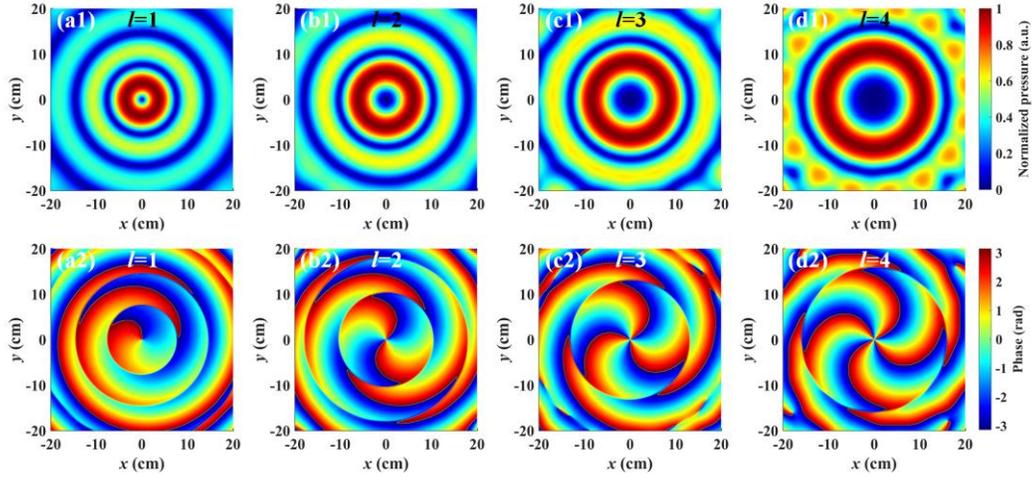

FIG. 2. (Color online) Simulated cross-sectional (a1)-(d1) pressure and (a2)-(d2) phase maps of AV beams with TC $l$= 1~4.

Subsequently, we calculated the corresponding $\omega_j^r$ and $\omega_j^i$ to precisely determine the specific TC to be extracted. Based on this approach, we separately extracted the modes of AV beams with $l$=1~4, and the outcomes of this extraction are presented in Table I. Table I's first and second rows present the $\omega_j^r$ and $\omega_j^i$ corresponding to the eigenvalues extracted from the mapping matrix $A$.

From the table, it can be observed that in the 4 times extraction process, the values of $\omega_j^r$ are consistently very close to zero. This alignment adheres to the condition discussed earlier in our document, indicating $\omega_j^i$ can be considered as a valid TC within the acoustic field. Consequently, as observed in the second row of Table I, the results of extracting the four TCs remain entirely consistent with the actual TCs contained within the acoustic field.

TABLE I. The DMD extraction results for AV beams with $l=1\sim4$.

|  | 1st | 2nd | 3rd | 4th |
| --- | --- | --- | --- | --- |
| $\omega_j^r$ | $1.74\times10^{-11}$ | $-1.41\times10^{-8}$ | $-1.34\times10^{-8}$ | $-3.21\times10^{-7}$ |
| $\omega_j^i$ | 1.00 | 2.00 | 3.00 | 4.00 |

Upon achieving accurate extraction of individual TCs, we further applied the DMD method to extract multiple distinct TCs from the multiplexed AV beams. Initially, we encoded the four English letters corresponding to Hefei University of Technology (HFUT) using the ASCII code table. Given our usage of AV beams with $l=-4\sim4$, we interpreted the on state as 1 and the off state as 0. The encoded information was transmitted in four times. The topological charges corresponding to the bits of the ASCII binary codes are presented in Table II. According to Table II, the data for the first transmission was 01001000B, corresponding to the coupling of two single AV beams with $l_1=-3$ and $l_2=1$. The data for the second transmission was 01000110B, indicating the coupling of three single AV beams with $l_1=-3$, $l_2=2$, and $l_3=3$, and so forth.

The pressure and phase distribution of the multiplexed AV beams for the transmission of the letters HFUT are shown in Fig.3. It is evident that the pressure distribution in the multiplexed AV beams no longer forms a circular ring, although it maintains a symmetrical distribution. Additionally, apart from the primary vortex at the center of the acoustic field, multiple off-axis sub-vortices[44] are also present, with their radial and axial positions varying with different transmitted letters. Since the acoustic field includes more than one TC, it becomes necessary to rearrange the received data matrix.

TABLE II. ASCII binary codes of the letters HFUT.

| Letter/$l$ | -4 | -3 | -2 | -1 | 1 | 2 | 3 | 4 |
|---|---|---|---|---|---|---|---|---|
| H | 0 | 1 | 0 | 0 | 1 | 0 | 0 | 0 |
| F | 0 | 1 | 0 | 0 | 0 | 1 | 1 | 0 |
| U | 0 | 1 | 0 | 1 | 0 | 1 | 0 | 1 |
| T | 0 | 1 | 0 | 1 | 0 | 1 | 0 | 0 |

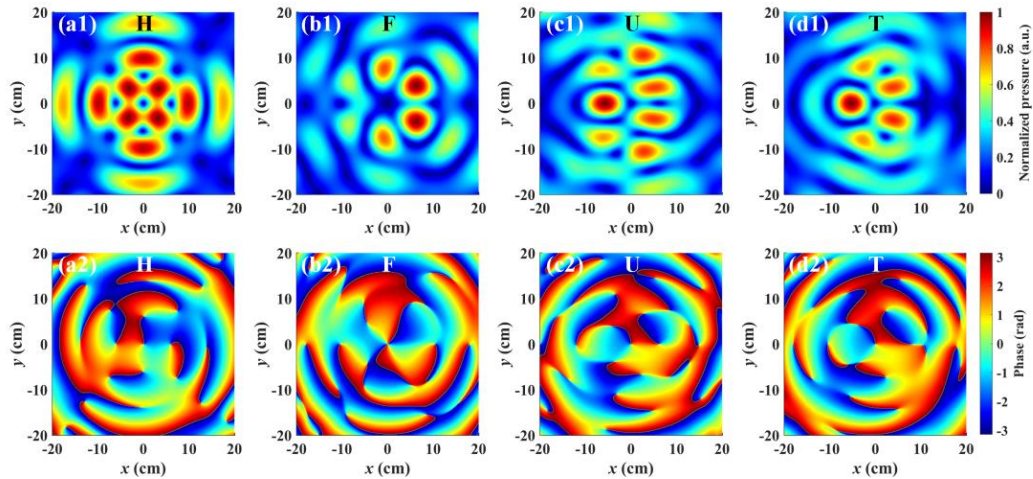

FIG. 3. (Color online) Simulated cross-sectional (a1)-(d1) pressure and (a2)-(d2) phase maps of multiplexed AV beams of transmitting the letters HFUT.

The specific value of the parameter $g$ needs to be adjusted, as it directly determines the number of TCs that can be extracted from the multiplexed AV beams. For the transmission of the letter H, it requires the utilization of two TCs and we set $g=2$. Subsequently, the received acoustic field data is rearranged into two data matrices, each sized at 2×14. Following this, DMD decomposition facilitates the extraction of two eigenvalues from the mapping matrix $A$. The corresponding values of $\omega_j^r$ and $\omega_j^i$ are then calculated to determine the specific TCs present in the acoustic field. We extracted the TCs corresponding to the multiplexed AV beams for the four transmitted letters through this approach. The decoding results for the four transmissions are summarized in Table III. The table reveals that the ASCII codes for the four transmitted letters are 01001000B, 01000110B, 01010101B, and 01010100B, which consistent with the ASCII code for the letters HFUT. This demonstrates the successful transmission of data underwater. As we employed partial circular arc-sampling, this receiving method significantly breaks the orthogonality between AV beams with different TCs. Traditional orthogonal decoding and OAM spectrum decomposition methods are no longer applicable. Hence, the DMD scheme proposed in this study does not rely on the orthogonality between AV beams with different TCs, making it particularly suitable for scenarios where a complete acoustic field distribution is unavailable and where the orthogonality between AV beams is severely compromised.

TABLE III. The DMD extraction results of multiplexed AV beams transmitting the letters HFUT.

|     | Decoding results | | | |
| --- | --- | --- | --- | --- |
| 1st | -3.00 | 1.00 | / | / |
| 2nd | -3.00 | 1.98 | 3.00 | / |
| 3rd | -3.00 | -1.04 | 2.00 | 3.99 |
| 4th | -3.00 | -1.00 | 2.00 | / |

## IV. DISCUSSION

### A. Comparison with previous methods

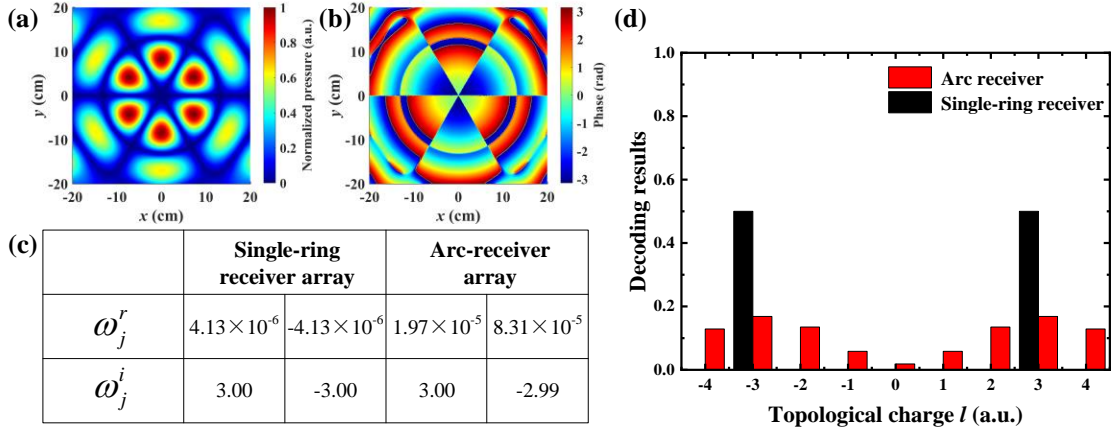

FIG. 4. (Color online) Simulated cross-sectional (a) pressure and (b) phase maps of multiplexed AV beams with TC $l=-3$ and $l=3$. The DMD decoding results (c) and orthogonal decoding results (d) of the single-ring receiver array and the arc-receiver array.

In this section, we conduct a comparative analysis by contrasting our modified DMD decoding scheme with the traditional orthogonal decoding approach. Initially, we employ a ring array to transmit AV beams with $l=-3$ and $l=3$. Figures. 4(a) and 4(b) depict the corresponding pressure and phase distributions for the multiplexed acoustic fields. Figure. 4(b) illustrates the TCs extracted by DMD from the acoustic field. It is observed that both sampling methods can accurately extract the TCs from the acoustic field. When using a ring array to collect data, the acoustic field information is more complete, resulting in slightly superior decoding results compared to the arc sampling scheme. As seen in Fig. 4(d), when the receiving array is a single-ring array, the decoding results are consistent with the actual TCs of $l=-3$ and $l=3$. However, when employing the arc sampling scheme, a considerable extension of the measured OAM spectrum is observed by using orthogonal decoding approach. Even with the use of a threshold, we cannot accurately determine the specific TC in the acoustic field. This

OAM spectrum extension arises from the breaking of orthogonality between AV beams with different TCs when employing the arc-receiving scheme. Consequently, incorrect OAM modes appear in the decoding results. Therefore, the DMD method proposed in this paper, which does not rely on the orthogonality between AV beams, is more suitable for demultiplexing acoustic fields in the arc sampling scheme.

### B. Performance analysis

In consideration of practical scenarios where the receiving array might not be perfectly aligned and various interferences exist in the communication system, it is essential to evaluate the performance of the DMD-based decoding scheme. Initially, we introduced Gaussian white noise with different signal to noise ratio (SNR) into the communication system. By transmitting the AV beam with topological charge $l=3$ for multiple times and the acoustic data were received by a 1/4 circular arc array consisting 16 elements. Subsequently, we applied OAM demodulation and statistically analyzed the total transmitted codes and the number of errors, calculating the bit error rate (BER) through multiple computations. Additionally, we simulated the decoding results of a traditional orthogonal decoding scheme based on the orthogonality between AV beams with the same transmitting and receiving elements and under identical interference conditions. The relationship between the BER and the SNR is depicted in Fig. 5, where we could observe that the overall BER gradually decreases with an increasing SNR. Under identical noise interference conditions, the DMD decoding scheme exhibits a significantly lower BER compared to the orthogonal decoding scheme. Furthermore, as the SNR gradually decreases, the orthogonality between AV beams is gradually disrupted. When the SNR drops to 17 dB, the BER exceeds 10%. However, even under such challenging conditions and even lower SNR, the DMD decoding scheme maintains a favorable BER. Therefore, in scenarios with severe noise interference in the communication environment, the DMD decoding scheme should be preferred.

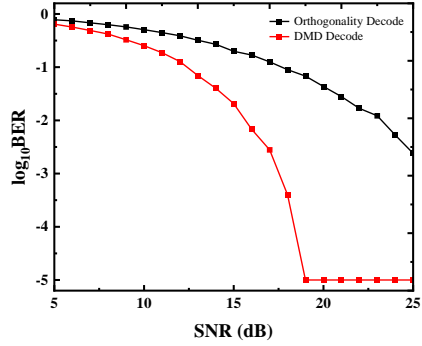

FIG. 5. (Color online) The variation of BER with respect to SNR for both decoding schemes.

Moreover, achieving precise alignment between the transmitting and receiving arrays in practical applications is exceptionally challenging. This challenge is particularly pronounced in communication systems based on vortex beams, which impose stringent alignment requirements on the transmitter-receiver arrays. Misalignment may lead to the degradation of communication performance. Therefore, researchers have also conducted corresponding analyses on the performance of systems under conditions where perfect alignment is not achieved, simultaneously proposing solutions to the alignment calibration issues for both transmitter and receiver arrays.[46] Here, it is equally crucial to assess the performance of the DMD decoding method under conditions of misalignment between the transmitting and receiving arrays. As depicted in Fig. 6(a), the diagram illustrates lateral misalignment between the receiving and transmitting arrays, where $\Delta y$ represents the displacement distance. To assess the impact of misalignment on BER, the receiving array is incrementally translated along the $y$-direction by 0.5 mm within a 10 mm range, under the conditions of SNR = 30 dB. BER for the system are computed at each position. The variation in BER for both decoding methods concerning displacement distance is presented in Fig. 6(b). It is evident that the relative positional changes of the transmitter and receiver arrays within a parallel plane can significantly impact the decoding results. The BER for both decoding methods increases notably with the displacement distance. However, it is clear that the orthogonal decoding results are more sensitive to misalignment. BER begins to steeply rise only when $\Delta y$ = 2.5 mm. When $\Delta y$ increases to 3.5 mm, the BER exceeds 10%. Hence, conventional orthogonal decoding schemes typically necessitate minimal offset between co-located transmitter and receiver arrays placed in parallel. In comparison to orthogonal decoding, it is apparent from the graph that the DMD decoding scheme exhibits a steep increase in BER only when the displacement distance reaches $\Delta y$ = 5 mm. Within the range of $\Delta y$ < 6 mm, DMD decoding scheme maintains a relatively low BER. The allowable range of misalignment between the transmitting and receiving arrays is significantly more extensive for DMD decoding scheme compared to orthogonal decoding scheme.

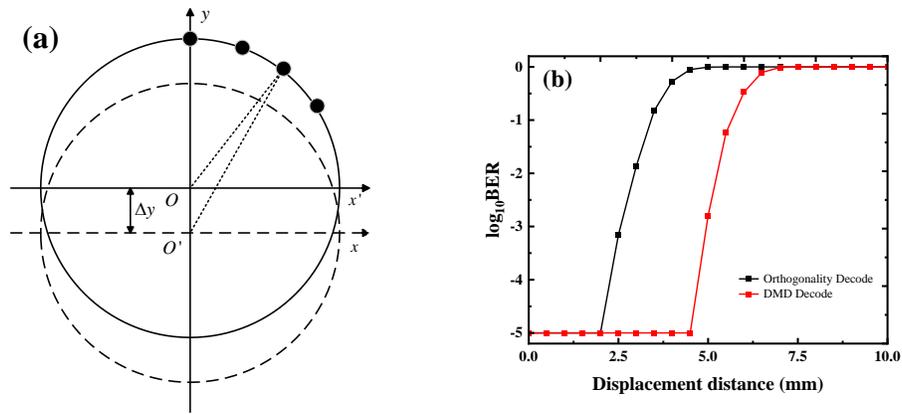

FIG. 6. (Color online) (a) Schematic of lateral misalignment in the receiving array. (b) Relationship between lateral displacement distance and BER.

In addition to misalignment between the receiving and transmitting arrays, practical experiments often encounter the challenge of misalignment due to the deflection of these arrays. As illustrated in the Fig. 7. (a), it represents a schematic of angular misalignment between the receiving array plane and the original receiving plane, where $\theta$ denotes the deflection angle. To analyze the impact of $\theta$ on the BER, under the SNR=30 dB condition, the receiving array is rotated along the original receiving plane in 0.1° increments for 20 iterations. The BER for both decoding schemes with respect to the deflection angle is shown in Fig. 7(b). It can be observed that even a slight deflection of the receiving array leads to a rapid increase in BER for the orthogonal decoding scheme, whereas the DMD decoding scheme remains largely unaffected within a certain range of deflection angles. When the deflection angle $\theta$ exceeds 0.2°, the BER for orthogonal decoding scheme already surpasses 10%, resulting in severe errors. In contrast, the DMD decoding scheme demonstrates a significant increase in BER only when the deflection angle reaches 1.1°. Furthermore, it maintains a low error rate for deflection angles below 1.3°. Considering various common system interferences observed in the experimental systems, we find that the performance of the DMD decoding scheme is significantly superior to traditional orthogonal decoding approaches. Therefore, the DMD scheme not only demonstrates a clear advantage in simplifying the design of the receiving array but also exhibits enhanced robustness and higher accuracy when dealing with potential external interferences encountered in practical applications.

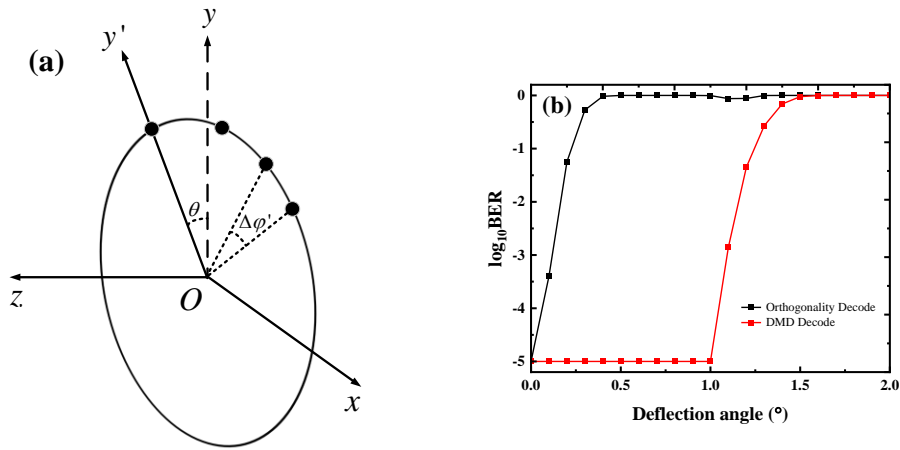

FIG 7. (Color online) (a) Schematic of angular misalignment in the receiving array. (b) Relationship between deflection angle and BER.

## V. CONCLUSION

In this paper, we illustrate an efficient AV beams demodulation scheme based on DMD. At the receiving end, we derive the principles of demodulation using a simplified circular arc-shaped receiving array based on an improved DMD approach. The communication system employs a circular arc-shaped receiving array composed of 16 receiving elements, in conjunction with a 16-element transmitting array. We encode the topological charge corresponding to the university abbreviation HFUT into multiple AV beams for transmission. The TCs within the acoustic field were successfully extracted by employing circular arc-shaped sampling in combination with the DMD method. Additionally, we compare the error performance of the DMD decoding scheme with the traditional orthogonal decoding scheme in response to common communication system interference. The results reveal that the DMD decoding scheme exhibits enhanced robustness. In summary, compared to traditional orthogonal-based decoding schemes, our scheme eliminates the need for complete acoustic field information acquisition and is independent of the orthogonality between AV beams. It enables the demodulation of multiplexed AV beams with only partial acoustic field information. Numerical simulation results indicate that the demodulation scheme based on DMD can effectively simplify the design of the receiving array and possess greater flexibility and adjustability, while offering increased robustness. This approach holds promise for widespread application in long-range underwater acoustic vortex communication.

## AUTHOR DECLARATIONS

### Conflict of Interest

The authors have no conflicts to disclose.

### DATA AVAILABILITY

The data that support the findings of this study are available from the corresponding author upon reasonable request.